\documentclass[11pt]{article}
\usepackage{graphicx}

\usepackage{amsmath,amssymb}
\usepackage{epsfig}
\usepackage{hyperref}
\usepackage{color}
\usepackage{cite}
\usepackage{multirow}

\newcommand{\hhref}[1]{\href{http://arxiv.org/abs/#1}{#1}}

\newcommand\pubnumber{}
\newcommand\pubdate{}

\def\institute{IFAE and BIST, Universitat Aut\`onoma de Barcelona,\\ E-08193 Bellaterra, Barcelona, Spain}
\def\support{\footnote{Work supported by the Spanish Ministry MEC under grant FPA2014-55613-P, by the Generalitat de Catalunya grant 2014-SGR-1450, by the Severo Ochoa excellence program of MINECO (grant SO-2012-0234 and SEV-2016-0588), and by the European Commission through the Marie Curie Career Integration Grant 631962 via the DESY-IFAE cooperation exchange.}}

\def\Title#1{\begin{center} {\Large #1 } \end{center}}
\def\Author#1{\begin{center}{ \sc #1} \end{center}}
\def\Address#1{\begin{center}{ \it #1} \end{center}}

\newcommand\pubblock{\rightline{\begin{tabular}{l} \pubnumber\\
         \pubdate  \end{tabular}}}
\newenvironment{Abstract}{\begin{quotation}  }{\end{quotation}}
\newenvironment{Presented}{\begin{quotation} \begin{center} 
             PRESENTED AT\end{center}\bigskip 
      \begin{center}\begin{large}}{\end{large}\end{center} \end{quotation}}

\def\beq{\begin{equation}}
\def\eeq#1{\label{#1}\end{equation}}
\def\eeqn{\end{equation}}

\def\beqa{\begin{eqnarray}}
\def\eeqa#1{\label{#1}\end{eqnarray}}
\def\eeqan{\end{eqnarray}}

\let\bar=\overbar

\def\Dslash{\not{\hbox{\kern-4pt $D$}}}
\def\dslash{\not{\hbox{\kern-2pt $\del$}}}

\def\msb{{\bar{\ssstyle M \kern -1pt S}}}

\begin{document}
\begin{titlepage}
\pubblock

\vfill
\Title{Top compositeness, Flavor and Naturalness}
\vfill
\Author{Giuliano Panico\support}
\Address{\institute}
\vfill
\begin{Abstract}
We discuss the connection between the top quark, naturalness and flavor in composite Higgs models.
The phenomenological features of the top quark and the associated fermionic partners are presented.
The main realizations of the flavor structure, in particular the anarchic partial compositeness scenario and
its possible modifications, are also briefly reviewed.
\end{Abstract}
\vfill
\begin{Presented}
$10^{th}$ International Workshop on Top Quark Physics\\
Braga, Portugal,  September 17--22, 2017
\end{Presented}
\vfill
\end{titlepage}
\def\thefootnote{\fnsymbol{footnote}}
\setcounter{footnote}{0}

\section{Introduction: Top Partners and Naturalness}

Together with the Higgs boson, the top quark is one of the main actors in beyond the Standard Model (SM) theories.
Its sizable Yukawa coupling, $y_{top}$, generates the largest radiative corrections to the Higgs potential. If these corrections
are too large, which is the case if the SM has a large cut-off $\Lambda_{\rm SM}$,
a sizable amount of cancellation is required to obtain the correct Higgs mass. This is the origin of the
well-known naturalness problem. To be more quantitative we can
evaluate the amount of fine-tuning by
comparing the physical Higgs mass $m_H \simeq 125\;$GeV with the leading one-loop corrections. In such a way we can estimate the
(minimal) amount of fine-tuning to be
\begin{equation}\label{eq:tuning}
\Delta \geq \frac{\delta_{\rm 1-loop} m_H^2}{m_H^2} = \frac{3 y_{top}^2}{8 \pi^2} \left(\frac{\Lambda_{\rm SM}}{m_H}\right)^2 \simeq \left(\frac{\Lambda_{\rm SM}}{450\;\rm{GeV}}\right)^2\,.
\end{equation}

It is clear that, in order to obtain a completely natural (i.e.~not fine tuned) theory, the
SM cut-off must be close to the electroweak (EW) scale. The UV dynamics, moreover, must be ``connected'' to the
top quark, in such a way to ``screen'' its radiative contributions to the Higgs potential. In such models the
$\Lambda_{\rm SM}$ scale can be typically identified with the mass of some ``partners'' of the top, which must
be below or around the TeV scale to avoid too much tuning.

This conclusion proves correct in basically all classical beyond the SM (BSM) theories that aim to solve the naturalness
problem. A well known example is low-energy supersymmetry, in which bosonic partners of the top, the `stops',
are present, which regulate the quadratic divergence in the radiative top contributions to the Higgs mass.

In the following we will focus on another class of theories, in which the Higgs boson is not an elementary
state as in the SM, but instead a composite object arising from a new
strongly-coupled dynamics. This idea reached nowadays a compelling embodiment, the ``composite
Higgs'' (CH) scenario (see~\cite{Contino:2010rs,Panico:2015jxa} for reviews).
Its main assumption is the identification of the Higgs with a pseudo-Goldstone
boson~\cite{ Kaplan:1983fs}, corresponding to an ${\rm SO}(5)/{\rm SO}(4)$ coset in minimal models~\cite{Agashe:2004rs}.

An additional ingredient is the generation of the top mass through the partial-compositeness
mechanism~\cite{Kaplan:1991dc} (we will discuss this point in detail in sec.~\ref{sec:flavor}).
Under this assumption the SM states, which are external with respect to the composite sector (with the
possible exception of the $t_R$ component), are mixed with suitable composite operators. In the low-energy description
these couplings can be effectively parametrized through mass mixings of the elementary states with composite fermionic
resonances. As can be easily understood, the resonances corresponding to the top quark, the `top partners',
play a central role in the EW dynamics. First of all they give rise to the top Yukawa coupling and, second, they control the generation
of the Higgs potential and therefore the amount of fine-tuning. Top partners constitute one of the most compelling prediction of
CH models and are one of the privileged ways to test directly these scenarios.

Higgs and top compositeness have other important implications for phenomenology. One of them is a quite specific
pettern of deviations in the Higgs and top couplings, which can be tested in collider experiments. Another interesting aspect
is the flavor structure, which is deeply influenced by partial compositeness. We will discuss all these aspects in the following sections.

\section{Phenomenology of Top Partners}\label{sec:top_partners}

Goldstone symmetry and partial compositeness determine the main properties of top partners.
Being part of the composite sector, they necessarily fill extended EW multiplets corresponding to representations of the
unbroken custodial group ${\rm SO}(4) \simeq {\rm SU}(2)_L \times {\rm SU}(2)_R$. Since they mix with the top quarks,
top partners must be charged under QCD transforming in the fundamental of ${\rm SO}(3)_c$.

As we mentioned the top and its partners give rise to the leading contribution to the Higgs potential through radiative effects.
The size of the Higgs mass term is controlled by the typical mass of the top partners (roughly coinciding with
the mass of the first complete ${\rm SO}(5)$ representation of states). Light top partners are thus needed to reduce the
fine tuning~\cite{Matsedonskyi:2012ym,Panico:2012uw}. The amount of tuning can be easily estimated through
eq.~(\ref{eq:tuning}) identifying $\Lambda_{\rm IR}$ with the top partners mass. Notice that this is a only a lower bound on
the fine-tuning, since in several models peculiarities of the structure of the Higgs potential require additional cancellations~\cite{Panico:2012uw}.

Top partners can be copiously produced at hadron colliders since they are colored objects. For low masses the
dominant production channel is QCD pair production, whose cross section only depends on the
resonance mass. This channel can thus be used to extract model-independent bounds on top partners. An additional production channel
is single production in association with a top or a bottom quark. This channel crucially depends on the EW gauge
couplings that mix the SM quarks with the composite partners and is thus
model-dependent~\cite{DeSimone:2012fs}.
It is more relevant for high partner masses, since its cross section
decreases more slowly than pair production.

\begin{figure}[t]
\centering
\includegraphics[width=.4\textwidth]{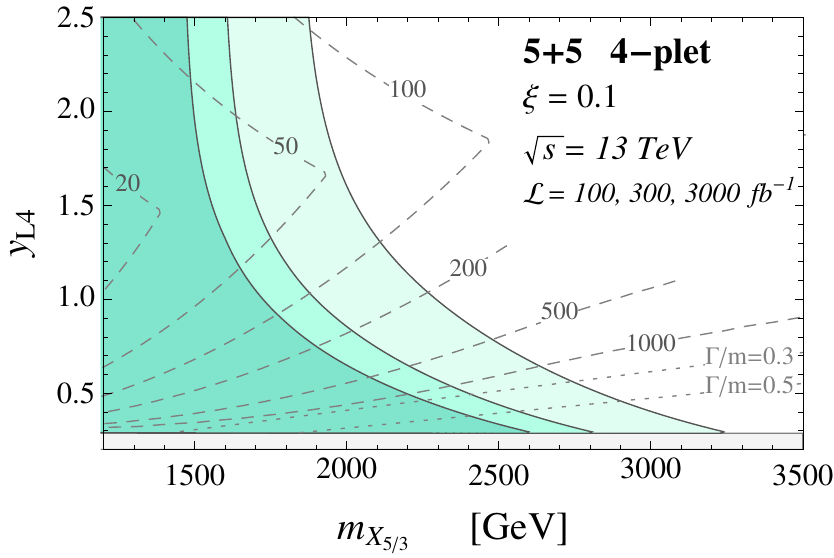}
\hspace{2em}
\includegraphics[width=.385\textwidth]{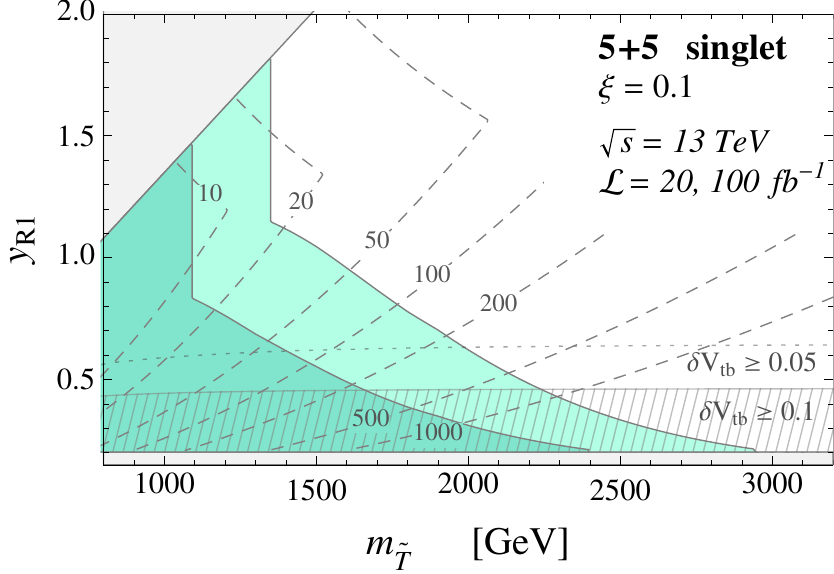}
\vspace{-1em}
\caption{Estimate of the LHC direct exclusion bounds on the top partner masses in two simplified effective scenarios with a light
${\rm SO}(4)$ fourplet (left) and a light singlet (right)~\cite{Matsedonskyi:2015dns}. The dashed lines show an estimate of the minimal amount of tuning $\Delta$.
The striped region in the right plot corresponds to the indirect bounds from deviations in the $V_{tb}$ matrix element.}
\label{fig:bounds}
\end{figure}

To give an idea of the LHC reach on top partners we show in fig.~\ref{fig:bounds} some projections for the bounds on typical top
partners multiplets at the $13\;$TeV LHC~\cite{Matsedonskyi:2015dns} (updated bounds can be found in~\cite{Panico:2017vlk}).
Current bounds from pair production are of order $1.2\;$TeV, basically independent of the
top parter quantum numbers. Single production bounds instead are more sensitive to the details of the model. In the plots we also
show the estimates of the minimal amount of fine-tuning obtained from eq.~(\ref{eq:tuning}). One can see that configurations with
relatively low tuning ($\sim 10\%$) are still allowed at present, whereas the end of the LHC program will test parameter space points
up to $\sim 1\%$ tuning.

In miminal CH scenarios, such as the holographic MCHM constructions~\cite{Agashe:2004rs}, the mass of the lightest top partners can be
directly related to the compositeness scale $f$, corresponding to the Goldstone Higgs decay constant~\cite{Matsedonskyi:2012ym}.
In this case the bounds from top partners direct searches can be translated into a lower bound on $f$. Current exclusions
correspond to $f \gtrsim 800\;$GeV (or $\xi = v^2/f^2 \lesssim 0.1$, where $v = 246\;$GeV is the Higgs VEV), while the end of the LHC program could push the bound
to $f \gtrsim 1.1\;$TeV ($\xi \lesssim 0.05$)~\cite{Matsedonskyi:2015dns}.

\section{Top Couplings}\label{sec:top_couplings}

The CH set-up gives rise to peculiar deviations in the Higgs and top couplings. The main effects are the modification of
Yukawa and gauge interactions and the presence of $4$-top effective operators.

All these effects are controlled by the Goldstone symmetry. The couplings to the gauge fields
are affected in a universal way. In models based on the ${\rm SO}(5) \rightarrow {\rm SO}(4)$ symmetry structure, the couplings to the
$W$ and $Z$ bosons are rescaled by a factor $\kappa_V = \sqrt{1 - \xi}$. The Yukawa couplings, on the other hand,
are modified in a way that depends on the quantum numbers of the fermion partners. Popular scenarios are the ones in which
the SM fermions are mixed with operators in the fundamental ${\rm SO}(5)$ representation, in which case the top Yukawa is rescaled
by a factor $\kappa_F = (1 - 2 \xi)/\sqrt{1-\xi}$, or in the spinorial representation, which gives $\kappa_F = \sqrt{1 - \xi}$.

The LHC measurement of the Higgs couplings can be used to derive robust bounds on the value of $\xi$. The current data
give a bound $\xi \lesssim 0.1$ (they are derived in scenarios in which the bottom and top Yukawa's are modified in the same
way, however similar bounds are typically found in more generic models)~\cite{Aad:2015pla}.
The present constraints are roughly a factor of two stronger than the expected ones, due to a positive shift in the
central value fit of $\kappa_V$. High-luminosity LHC data are not expected to significantly improve the bounds if the central value
will move closer to the SM prediction, whereas the bound could change substantially if some deviation will persist.

It is interesting to notice that values $\xi \simeq 0.1$ allow for sizable deviations in the top Yukawa. In models with composite
operators in the fundamental ${\rm SO}(5)$ representation corrections $\delta y_{top} \simeq 15 - 20\%$ are still compatible with the
experimental data. Such deviations could be tested by the direct determination of the top Yukawa in Higgs associated production
with a $t\bar t$ pair.

Other interesting modifications arise for the gauge interactions involving the top quark. Viable models typically
require a discrete custodial $P_{LR}$ symmetry to keep under control the $Z$ couplings to the bottom field~\cite{Agashe:2006at},
which are tested at the $0.1\%$ level. This symmetry, however, can not protect at the same time the top interactions.
Deviations of order $\xi$ are thus generically present, induced both by the mixing if the top with its partners and by the
presence of vector resonances that mix with the SM gauge fields. Deviations in the $Z \bar t_L t_L$ coupling are constrained
by the EW precision measurement to be $|\delta g_{Zt_L}| \lesssim 8\%$~\cite{Efrati:2015eaa}. The $Zt_R\overline t_R$ interaction,
on the other hand, is much more elusive, since its SM value is suppressed, and is basically unconstrained.

Another coupling that can receive sizable modifications is $W \overline t_L b_L$, which corresponds to the $V_{tb}$ CKM element.
In minimal models with custodial $P_{LR}$, deviations in this coupling are related to the corrections in the $Z \bar t_L t_L$ vertex,
namely $\delta g_{Zt_L} = \delta V_{tb}$~\cite{delAguila:2000rc}. The bounds on $\delta V_{tb}$ can imply non-trivial constraints on
the top partners parameter space, which can be competitive with direct searches for heavy resonances (see for instance the right panel
of fig.~\ref{fig:bounds}).

Finally the sizable mixing of the top with its composite partners gives rise to effective $4$-fermion contact interactions
of the type ${\cal O} = c/f^2 (\bar t \gamma^\mu t)^2$. The coefficient $c$ parametrizes the amount of mixing of the top with its
partners and takes values $c \sim 1$ for sizable top compositeness. Contact $4$-top interactions can be tested in $\bar t t \bar t t$
production at the LHC. The currect bounds are of order $c/f^2 \lesssim 1/(590\;{\rm GeV})^2$~\cite{ATLAS:2016btu}.

\section{Flavor Structure}\label{sec:flavor}

Let's now discuss the implications for flavor physics. Contrary to models with an elementary Higgs, in which
the Yukawa structure can originate in the far UV, in CH models the origin of flavor must be addressed
at much lower energies. Since the Higgs is associated with a composite operator
${\cal O}_H$ whose dimension must satisfy $d_H > 2$ to avoid the hierarchy problem, the Yukawa interactions
$\overline f_L {\cal O}_H f_R$ have dimension larger than $4$ and are irrelevant operators. If $\bar f_L {\cal O}_H f_R$ is generated
at very high energies fermion masses are necessarily too small. To be more quantitative,
if $d_H \gtrsim 2$, the maximal energy scale at which a realistic top Yukawa can be generated is $\Lambda_t \lesssim 10\,$TeV.

The classical approach to flavor in composite Higgs models is based on the partial compositeness idea,
in which the SM fermions get masses by mixing linearly with strong sector operators
${\cal L}_{\rm lin} = \varepsilon_{f_i} \overline f_i {\cal O}_i$.
At the strong scale, $\Lambda_{\rm IR} \sim \rm{TeV}$, the fermion Yukawa's are generated with a pattern ${\cal Y}_f \sim g_* \varepsilon_{f_i} \varepsilon_{f_j}$, where $g_*$ is the typical strong-sector coupling. The appealing feature of this scenario,
usually dubbed ``anarchic partial compositeness''~\cite{Kaplan:1991dc,anarchic}, is the fact that the smallness of the mixings $\varepsilon_{f_i}$ can
simultaneously explain the smallness of the fermion masses and mixing angles.

This set-up, however, also predicts sizable
flavor-violating effects. Large contributions are expected for the neutron EDM and for $\epsilon_K$, which force the
$\Lambda_{\rm IR}$ scale to be of order $10\;$TeV requiring a significant amount of tuning~\cite{anarchic_bounds,Panico:2015jxa}.
The situation is even worse if anarchic partial compositeness is naively extended to the lepton sector, in which case corrections to the
electron EDM and large $\mu \rightarrow e \gamma$ transitions require $\Lambda_{\rm IR} \gtrsim 100\;$TeV~\cite{Panico:2015jxa}.

A possible way to avoid large flavor effects is to introduce flavor symmetries, assuming that the right-handed quarks are fully-composite
objects~\cite{Redi:2011zi}. Although these models can pass the flavor bounds for $\Lambda_{\rm IR} \sim {\rm TeV}$, they predict sizable deviations
in dijet distributions and large production cross sections for multi-TeV resonances that translate in stringent collider
bounds~\cite{Domenech:2012ai}.

A recently proposed departure from the classical anarchic paradigm can be used to avoid severe flavor and CP-violating
constraints~\cite{Vecchi:2012fv,Panico:2016ull}.
The main idea is to assume that the operators ${\cal O}_{f_i}$, that mediate the mixing between the SM fermions and the Higgs, decouple
at different energy scales $\Lambda_{f_i} \gg \Lambda_{\rm IR} \sim {\rm TeV}$. This implies that Yukawa-like couplings
$\overline f_i {\cal O}_H f_j$ are generated at scales larger than $\Lambda_{\rm IR}$, avoiding sizable flavor and CP-violating effects.
The hierarchies in the fermion spectrum and the smallness of the mixing angles is now explained in a ``dynamical'' way by the different $\Lambda_{f_i}$ scales: the larger the decoupling scale, the smaller the Yukawa coupling.
In this set-up the only Yukawa that needs to be generated at a low scale is the top one, so that the usual partial compositeness
structure will still be valid for the top sector. All the other Yukawa's can be generated at significantly higher scales, up to
$\sim 10^7\;$TeV for the first-generation fermions.

The structure of the Yukawa matrix for the down quark sector is approximately
\begin{equation}
{\cal Y}_{\rm down}\simeq
\left(
\begin{array}{ccccc}
\ \ \  Y_d && \alpha_R^{ds} Y_d && \alpha_R^{db} Y_d\\
\rule{0pt}{1.25em}
 \alpha_L^{ds} Y_d &&  \ \ \ Y_s && \alpha_R^{sb} Y_s
\\
\rule{0pt}{1.25em}
 \alpha_L^{db} Y_d &&   \alpha_L^{sb} Y_s && \ \ \ Y_b
\end{array}
\right)\,,
\label{matrixYd}
\end{equation}
where $Y_i$ denote the SM Yukawa's and $\alpha_{L,R}$ are numerical coefficients typically of order one. An analogous formula
holds for the up sector. The off-diagonal terms are much smaller than in the anarchic case (in particular for the
up sector), leading to a suppression of flavor-changing effects.

In this scenario the leading flavor and CP-violating effects arise from the top partial compositeness~\cite{Panico:2016ull}.
Additional contributions originating at the decoupling scale of the other SM fermions are instead well below the current bounds.
The leading effects come from the two operators
\begin{equation}
(1/\Lambda_{\rm IR})\, (\overline Q_{L3} \gamma^\mu Q_{L3})^2\,,
\qquad (g_*/\Lambda_{\rm IR}) \overline Q_{L3} \gamma^\mu Q_{L3} i H^\dagger \overleftrightarrow{D}_\mu H\,.
\end{equation}
After the rotation to the mass eigenstate basis, these two operators give rise to $\Delta F = 2$ and $\Delta F = 1$ transitions
respectively. An interesting consequence is the fact that these corrections automatically follow a minimal flavor violation structure,
thus significantly ameliorating the compatibility with the experimental data. The current constraints (in particular from
$\varepsilon_K$, $\Delta M_B$, $\varepsilon'/\varepsilon$, $K \rightarrow \mu\mu$ and $B \rightarrow (X) \ell \ell$) can be satisfied for $\Lambda_{\rm IR} \sim \textit{few}\;$TeV.

Remarkably, the extension of this construction to the lepton sector avoids large corrections to the electron electric dipole moment (EDM)
and $\mu \rightarrow e \gamma$, relaxing the strong bounds of anarchic models to the level
$\Lambda_{\rm IR} \sim \textit{few}\;$TeV~\cite{Panico:2016ull}. The leading bounds arise from two-loop Barr--Zee contributions
to the electron EDM involving a loop of the top and its partners. The present experimental bounds allow to test
partner masses of order $\textit{few}\;$TeV, whereas near future experiments will improve the reach by more than one order
of magnitude, testing the natural parameter space of these models~\cite{Panico:2017vlk}.

\end{document}